%% file: cconnect.tex
\RequirePackage{xcolor}
\documentclass[journal,twoside,web]{ieeecolor}
\input{./include/preamble}

\input{./include/acronyms}

\newcommand{\argmin}{\operatornamewithlimits{arg\,min}}

\def\BibTeX{{\rm B\kern-.05em{\sc i\kern-.025em b}\kern-.08em
		T\kern-.1667em\lower.7ex\hbox{E}\kern-.125emX}}

\begin{document}
\title{CConnect:  Synergistic Convolutional Regularization for Cartesian $T_2^{*}$ Mapping}
\author{Juan Molina, Alexandre Bousse,  Tabita Catal\'an, Zhihan Wang, Mircea Petrache, Francisco Sahli, Claudia Prieto and Mat\'ias Courdurier
\thanks{This work was supported  by the Millennium Nucleus ACIP NCN19-161, the Millenium Institute iHEALTH ICN2021-004, ANID Fondecyt projects 1210637, 1210426, 1240200, 1191903, and  the French National Research Agency (ANR) under grant No ANR-20-CE45-0020.}
\thanks{Juan Molina and Matias Courdurier are with the Department of Mathematics, Pontificia Universidad Cat\'olica de Chile, Santiago, Chile, and with the Millennium Nucleus for Applied Control and Inverse Problems, Santiago, Chile.}
\thanks{Alexandre Bousse and Zhihan Wang are with the Univ. Brest, LaTIM, Inserm, U1101, 29238~Brest, France.}
\thanks{Tabita Catal\'an is with the Millennium Nucleus for Applied Control and Inverse Problems, Santiago, Chile, and with the Millennium Institute for Intelligent Healthcare Engineering, Santiago, Chile.}
\thanks{Mircea Petrache is with the Department of Mathematics, Pontificia Universidad Cat\'olica de Chile, Santiago, Chile.}
\thanks{Francisco Sahli and Claudia Prieto are with the School of Engineering, Pontificia Universidad Cat\'olica de Chile, Santiago, Chile, with the Millennium Institute for Intelligent Healthcare Engineering, Santiago, Chile and the Millennium Nucleus for Applied Control and Inverse Problems, Santiago, Chile.}
\thanks{Corresponding author: Juan Molina, \texttt{jjmolina@mat.uc.cl}}
}

\maketitle
 
\input{./content/abstract}

\acresetall


\input{./content/introduction}

\input{./content/problem}
\input{./content/method}
\input{./content/experiments}
\input{./content/discussion}
\input{./content/conclusion}

\section*{Acknowledgments}

All authors declare that they have no known conflicts of interest in terms of competing financial interests or personal relationships that could have an influence or are relevant to the work reported in this paper.

We would like to thank the UNIACT team at Neurospin, CEA for the data acquisition.

\bibliographystyle{ieeetr}
{\footnotesize\bibliography{references}}

\end{document}

%% file: include/preamble.tex
\usepackage{generic}
\usepackage{textcomp}
\usepackage{mathtools}
\usepackage{graphicx}
\usepackage[export]{adjustbox}
\usepackage{float}
\usepackage{hyperref}
\usepackage{acro}
\usepackage{placeins}
\usepackage{overpic}
\usepackage{pgfplots}
\usepackage{subfigure}
\usepackage{caption}
\usepackage{cite}
\usepackage{color}

\usepackage{amsmath,amssymb,amsfonts}
\usepackage{algorithmic}

\usepackage{textcomp}
\def\BibTeX{{\rm B\kern-.05em{\sc i\kern-.025em b}\kern-.08em
    T\kern-.1667em\lower.7ex\hbox{E}\kern-.125emX}}

%% file: include/acronyms.tex
\DeclareAcronym{2D}{
	short=2-D,
	long=two-dimensional,
}

\DeclareAcronym{3D}{
	short=3-D,
	long=three-dimensional,
}

\DeclareAcronym{PET}{
	short=PET, 
	long=positron emission tomography,
}

\DeclareAcronym{MRI}{
	short=MRI, 
	long=magnetic resonance imaging,
}

\DeclareAcronym{MBIR}{
	short=MBIR, 
	long=model-based iterative reconstruction,
}

\DeclareAcronym{TV}{
	short=TV, 
	long=total variation,
}

\DeclareAcronym{JTV}{
	short=JTV, 
	long=joint total variation,
}

\DeclareAcronym{CNN}{
	short=CNN, 
	long=convolutional neural network,
}

\DeclareAcronym{PSNR}{
	short=PSNR,
	long=peak signal-to-noise ratio,
}

\DeclareAcronym{SSIM}{
	short=SSIM,
	long=structural similarity index measure,
}

\DeclareAcronym{ML}{
	short=ML,
	long=machine learning,
}

\DeclareAcronym{DL}{
	short=DL,
	long=deep learning,
}

\DeclareAcronym{qMRI}{
	short=qMRI ,
	long=quantitative MRI,
}

\DeclareAcronym{LR}{
	short=NLR ,
	long=nuclear low-rank,
}

\DeclareAcronym{MR}{
	short=MR ,
	long=magnetic resonance,
}

\DeclareAcronym{WR}{
	short=WR ,
	long=\ensuremath{\ell}\textsubscript{1}-wavelet regularization,
}

\DeclareAcronym{WT}{
	short=WT ,
	long=wavelet transform,
}

\DeclareAcronym{GT}{
	short=GT,
	long=ground truth,
}

\DeclareAcronym{DeAli}{
	short=DeAli ,
	long=de-aliasing,
}

\DeclareAcronym{DB}{
	short=DB ,
	long=dictionary learning,
}

\DeclareAcronym{DiL}{
	short=DiL ,
	long=dictionary learning,
}

\DeclareAcronym{IR}{
	short=IR ,
	long=iterative reconstruction,
}

\DeclareAcronym{CS}{
	short=CS ,
	long=compressed sensing,
}

\DeclareAcronym{TNV}{
	short=TNV,
	long=total nuclear variation,
}

\DeclareAcronym{PLS}{
	short=PLS,
	long=parallel level-set,  
}

\DeclareAcronym{CT}{
	short=CT,
	long=computerized tomography,  
}

\DeclareAcronym{T2w}{
	short=$T_{2}^*w$,
	long=$T_{2}^*$-weighted,  
}

\DeclareAcronym{CG}{
	short=CG,
	long=conjugate gradient,  
}

\DeclareAcronym{ADMM}{
	short=ADMM,
	long=alternating direction method of multipliers,  
}

\DeclareAcronym{RF}{
	short=RF,
	long=radio-frequency,  
}

%% file: content/abstract.tex
\begin{abstract}
    Magnetic resonance imaging (MRI) is fundamental for the assessment of many diseases, due to its excellent tissue contrast characterization. This is based on quantitative techniques, such as $T_1$, $T_2$, and $T_2^*$ mapping. Quantitative MRI requires the acquisition of several contrast-weighed images followed by a fitting to an exponential model or dictionary matching, which results in undesirably long acquisition times. Undersampling reconstruction techniques are commonly employed to speed up the scan, with the drawback of introducing aliasing artifacts. However, most undersampling reconstruction techniques require long computational times or do not exploit redundancies across the different contrast-weighted images. This study introduces a new regularization technique to overcome aliasing artifacts, namely CConnect, which uses an innovative regularization term that leverages several trained \acp{CNN} to connect and exploit information across image contrasts in a latent space. We validate our method using in-vivo $T_2^*$ mapping of the brain, with retrospective undersampling factors of 4, 5 and 6, demonstrating its effectiveness in improving reconstruction in comparison to state-of-the-art techniques. Comparisons against joint total variation, nuclear low rank and a \ac{DL} de-aliasing post-processing method, with respect to \ac{SSIM} and \ac{PSNR} metrics are presented. 
\end{abstract}

%% file: content/introduction.tex
\section{Introduction}

Quantitative \ac{MRI} has emerged in recent years as an important non-invasive tool to evaluate a range of neurological conditions, showing great potential for objective characterization of different tissue properties.  \cite{MRIbasic-hearth,qMRI-basic}. 

\Ac{qMRI} involves extracting measurable properties to assess the tissue and organs of interest. This includes for example $T_1, T_2$ and $T_2^*$ mappings \cite{qMRI-basic2}. Several clinical studies have shown the potential of $T_1$ mapping to characterize multiple sclerosis lesions or tumors \cite{T1-sclerosis,T1-tumor}. $T_2$ maps have been shown to be valuable in the brain detecting abnormalities in focal epilepsy and cognitive decline \cite{T2norm-epilepsy,T2normal-brain-ageing}. Whereas $T_{2}^{*}$ mapping is influenced by iron and myelin content and has been shown to be relevant in characterizing stroke and investigating neurodegenerative diseases, among others \cite{T2*neurodisease, T2*ironcontent, T2*stroke}. 

Conventionally, \ac{qMRI} parametric mapping requires the acquisition of several \ac{3D} high-resolution contrast-weighed images followed by a fitting to an exponential model or dictionary matching to generate the corresponding map. Consequently, acquiring  \ac{qMRI} data demands a considerable amount of acquisition time. In order to mitigate this challenge, a common approach is to acquire undersampled measurements \cite{subsampling}. Reconstructing the weighted-contrast images from undersampled data is an ill-posed inverse problem, and regularized undersampling reconstruction techniques are needed to minimize aliasing artifacts \cite{ill-posed-badquality}. 

Several iterative regularized reconstruction techniques have been proposed to speed up MRI acquisition \cite{MRI-iterative,mribasic-CSmethods}. These methods involve iteratively minimizing a functional composed of two components: the data fidelity term, which measures the discrepancy between predictions and acquired data, and the regularization (or penalty) term, which incorporates prior information about the solution. Typically, the fidelity term employs a squared error, while the regularization term has seen numerous innovations over time \cite{mribasic-CSmethods, CS-MRI-1}. Initial regularization efforts emphasized imposing sparsity of the solution using a sparsifying transform combined with a sparsity-promoting norm, such as \ac{TV} and $l_1$-wavelet regularization \cite{block2007undersampled,lustig2007sparse,lustig2008compressed}. However, these initial methodologies tended to overlook the redundancies between contrast-weighted images, which is especially relevant in parametric mapping techniques.

A second category of regularization techniques aims at promoting structural similarity between the image contrasts. Examples in this category include \ac{JTV} \cite{JTVcontrasts} and \ac{LR}  \cite{ mribasic-CSmethods, CS-MRI-1}. \ac{JTV} promotes similar contours between the contrasts, while \ac{LR} is designed to exploit redundant local information across contrasts. Despite their effectiveness,  the \ac{JTV} and \ac{LR} regularizations are custom-built, they do not always reflect the inherent characteristics of organ-specific anatomic images and may over-regularize pathology signals. Hence, there is still a need to improve \ac{MRI} image reconstruction techniques to reduce the scanning time and enhance the overall quality of the reconstructed images \cite{MRI-ML-BART, Learning-cartesian-ML}.

Recent years have been marked by the emergence of \ac{ML} approaches to solve inverse problems \cite{ML-InverseProblemgeneral,arridge2019solving,ML-InverseProblems}. In image reconstruction, non-deep \ac{ML} include \ac{DiL}, which has been used in \ac{MRI} \cite{caballero2014dictionary,ravishankar2010mr} as well as in \ac{CT} \cite{xu2012low}. More recently, \ac{DL} techniques have been widely investigated in medical image reconstruction  in \ac{CT}, \ac{PET} and \ac{MRI} \cite{bousse2024review,reader2020deep,wang2020deep}. Such approaches include, for example, (i) unrolling architectures, which consist of mimicking an iterative reconstruction algorithm with a \ac{DL} architecture \cite{monga2021algorithm}, (ii) direct reconstruction architectures (i.e., raw data to image) \cite{li2019learning,zhu2018image} and (iii) penalized iterative reconstruction with penalties learned via \ac{DL} architectures \cite{biswas2019dynamic, ML-InverseProblems,wang2016accelerating,pinton2024multi}.

Most of the above-mentioned techniques are limited to single-contrast reconstruction. They could benefit to be extended to multicontrast reconstruction in order to take advantage of the information shared across different weighted-contrast images, which is the case in multiparametric mapping. In this paper, we introduce an innovative iterative reconstruction framework that takes advantage of multiple trained \acp{CNN} for a penalized iterative reconstruction with a learned penalty. The proposed approach is investigated in $T_2^*$ mapping of the brain with retrospective undersampled Cartesian acquisition. Inspired from \cite{Alexpaper}, our novel synergistic regularization term leverages redundant information across various image contrasts by the mean of a collection of image-to-image \acp{CNN} which connect each contrast to a reference image. Besides, our \acp{CNN} are specifically designed and trained to systematically eliminate aliasing artifacts induced by the Cartesian undersampling. This approach, namely ColorConnect, seeks to enhance the quality and fidelity of the reconstructed \ac{qMRI} images.

The paper is structured as follows: Section~\ref{sec:problem} addresses the inverse problem associated with undersampled multi-contrast image reconstruction. Section~\ref{sec:method} describes how to obtain the $T_2^*$ map from the multi-contrast images and offers a concise review of the state-of-the-art of iterative reconstruction techniques for multi-contrast reconstruction. It also outlines classic methods and standard ML techniques, and presents our proposed approach as well as discussing potential alternatives. Section~\ref{sec:experiments} introduces in-vivo experiments, presenting results for $T_{2}^{*}$ weighted-contrast images and the corresponging $T_{2}^{*}$ map, in comparison to state-of-the-art reconstruction methods. Finally, Section~\ref{sec:discussion} discusses our results and the limitations of the proposed approach, while Section~\ref{sec:conclusion} summarizes the conclusion.

%% file: content/problem.tex
\section{Undersampled Multi-contrast Problem Formulation}\label{sec:problem}

In this work, we consider \ac{2D} images although it can be extended to \ac{3D} volumes without major changes.

\subsection{Forward \ac{MR} model}\label{sec:forward}

The \ac{MR} signal at time $t>0$ and spatial frequency $\vec{k}$ is given by the function $S$ defined as \cite{MRI-iterative, ForwardOperator}
\begin{align}
    S\left(t,\vec{k}\right)  {} = {} &  \int c(\vec{r})M(\vec{r},t)\mathrm{e}^{-2\iota\pi\vec{k}\cdot\vec{r}}\, \mathrm{d}\vec{r} \label{eq:ForwardSignal} \\
    						 {} = {} &  \mathcal{F} [c(\cdot) M(\cdot,t)]  (\vec{k})\label{eq:ForwardSignal2} & 
\end{align}
where $M(\vec{r},t)$ is the magnetization at location  $\vec{r}\in\mathbb{R}^{2}$ and time $t$,  $c$ is the coil sensitivity,  $\iota=\sqrt{-1}$ and $\mathcal{F}$ denotes the \ac{2D} spatial Fourier transform. 

For $T_{2}^{*}$ map, the magnetization is assumed to be given by 
\begin{equation}
    M(\vec{r},t)=M_{0}(\vec{r})\mathrm{e}^{-\iota\gamma B_{0}(\vec{r})t}\mathrm{e}^{-t/T_{2}^{*}(\vec{r})}      \label{eq:ModelT2}
\end{equation}
where $B_{0}$ represents the static magnetic field, $M_{0}$ is the initial net magnetization, $T_{2}^{*}$ is the transversal relaxation map and $\gamma$ is a constant  (approximately 42.58 MHz/T for hydrogen). 
%Therefore, the signal $S$ is the Fourier transformation of the coil response times the magnetization \cite{MRI-iterative, ForwardOperator}. 
For simplicity, we assume a single coil acquisition, i.e., $c$  is constant and equal to one.

Let $\mathcal{K}=\{\mathbf{k}_l\}_{l=1}^N\subset \mathbb{R}^2$ be a finite uniform rectangular grid in the k-space, with $N$ nodes. In the following, given a mapping $g:\mathbb{R}^2\to\mathbb{C}$, $g(\mathbf{k})\in \mathbb{C}^N$ denotes the $N$-dimensional vector defined at each node $l\in\{1,\dots,N\}$ as $[g(\mathbf{k})]_l=g(\mathbf{k}_l)$.

Let $t_i$ be the acquisition time corresponding to the $i$th time frame, with $i=1,\dots,n_\mathrm{t}$. For each $i$, let $U_i \subset \{1,\dots,N\}$, $|U_i| = N_i\leq N$,  be a subset of indices such that $\{\mathbf{k}_l\}_{l\in U_i}$ is the subset of $\mathcal{K}$ used to sample the acquisition during frame $i$.

For each $i$ define $\mathcal{P}_i:\mathbb{C}^N\to \mathbb{C}^{N_i}$ the subsampling operator given by 
\begin{align}
    \mathcal{P}_i(\{z_l\}_{l=1}^N) = \{z_l\}_{l\in U_i}, \quad \forall \{z_l\}_{l=1}^N\in\mathbb{C}^N.
\end{align}
 
Let $Y_i \in \mathbb{C}^{N_i}$ represent the $i$th measurement in the k-space. Following the forward model introduced above in this section (with $c=1$),  we have
\begin{align}
    Y_i  {} & {} =   \mathcal{P}_i\left( S \left(t_i,\mathbf{k}\right)\right) + \epsilon_i \nonumber \\
    {} & {} = \mathcal{P}_i\left( \mathcal{F} [ M(\cdot,t_i)]  (\mathbf{k})\right) +\epsilon_i \label{eq:ContinousOperator} 
\end{align}
where $\epsilon_i \in \mathbb{C}^{N_i}$ is a zero-mean measurement noise. 

\subsection{Inverse problem}

The inverse problem is first formulated as
\begin{align}
    & \text{find} \quad M(\cdot,t_i) \quad  \text{s.t.} \nonumber\\   
    & Y_i \approx  \mathcal{P}_i\left( \mathcal{F} \left[ M(\cdot,t_i)\right]  (\mathbf{k})\right)  \quad \forall i\label{eq:invprob} 
\end{align}
Equation~\eqref{eq:invprob} is a semi-discrete inverse problem and needs to be fully discretized for numerical experiments.

\begin{figure}[htbp]
	\centering
	\begin{overpic}[width=\linewidth, tics=5]{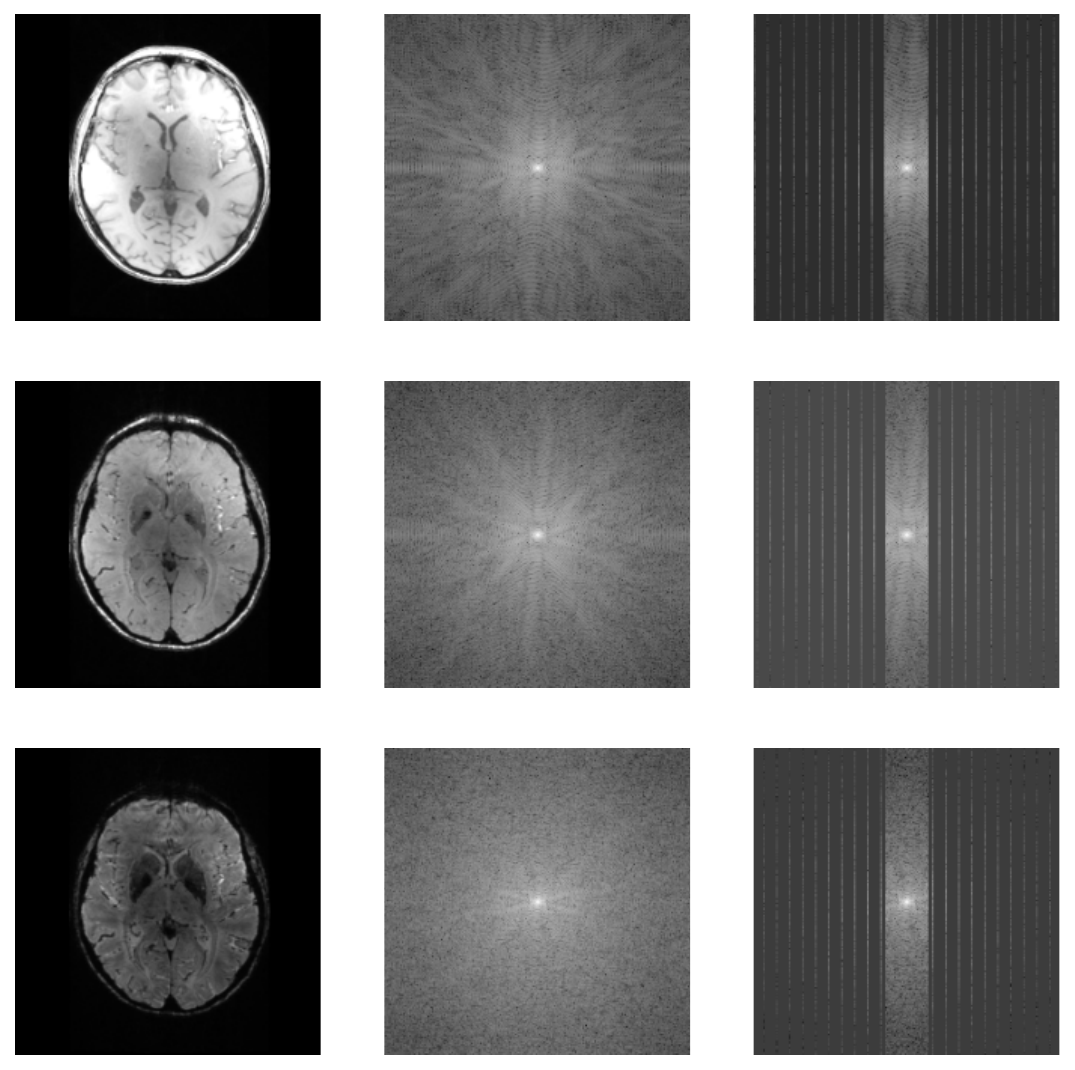}
		\put (30.3,51) {$\mathcal{F}^\mathrm{d}$}
		\put (30.7,47) {$\Rightarrow$}
		\put (65,51) {$\mathcal{P}_i$}
		\put (65,47) {$\Rightarrow$}
		
		\put (30.3,17) {$\mathcal{F}^\mathrm{d}$}
		\put (30.7,13) {$\Rightarrow$}
		\put (65,17) {$\mathcal{P}_i$}
		\put (65,13) {$\Rightarrow$}
		
		\put (30.3,85) {$\mathcal{F}^\mathrm{d}$}
		\put (30.7,81) {$\Rightarrow$}
		\put (65,85) {$\mathcal{P}_i$}
		\put (65,81) {$\Rightarrow$}
			
	\end{overpic}
	\caption{Illustration of the forward operator  $\mathcal{A}_{i}=\mathcal{P}_{i}\mathcal{F}^\mathrm{d}$ applied to three different $X_i$. The k-space images are represented by their complex magnitude.}\label{fig:Operator}
\end{figure}
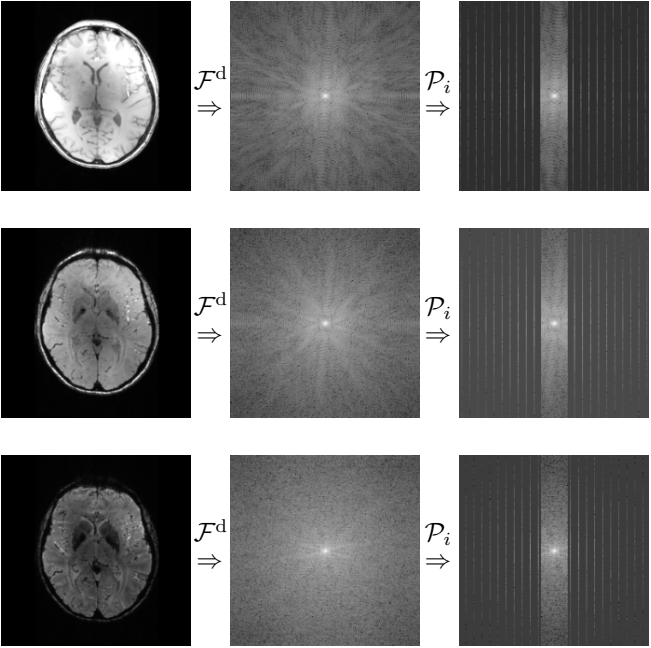

Assume $M(\cdot,t)$ lies in a finite-dimensional space defined by $N$ pixels such that
\begin{equation}
    M(\vec{r},t)=\sum_{j=1}^{N}x_j(t) m_j(\vec{r})\label{eq:BasisApproximation} 
\end{equation}
where each $m_j \colon \mathbb{R}^2 \to \mathbb{R}$ is a function spatially describing the $j$th pixel center at position $r_j\in\mathbb{R}^2$ (we require that $m_j(r_i)=\delta_{ij}$, the Kronecker delta) and $x_j(t)\in\mathbb{C}$ is the linear factor of pixel $j$ at time $t$. Substituting (\ref{eq:BasisApproximation}) in (\ref{eq:ContinousOperator}) and using the linearity of the operators, we have
\begin{align} 
    \mathcal{P}_i\left(\mathcal{F} [ M(\cdot,t_i)]  (\mathbf{k}\right))\nonumber & =\sum_{j=1}^{N}\mathcal{P}_i\left(\mathcal{F}\left[m_j \right](\mathbf{k})\right) x_j(t_i)\nonumber \\
    & = \mathcal{A}_{i}X_i
\end{align}
where $X_i=\left(x_{1}(t_i), \dots, x_{N}(t_i)\right)^T\in \mathbb{R}^N$, and $\mathcal{A}_{i}=\mathcal{P}_{i}\mathcal{F}^\mathrm{d}$ with $\mathcal{F}^\mathrm{d}\in\mathbb{C}^{N\times N}$ defined as $\left[\mathcal{F}^\mathrm{d}\right]_{l,j} = \mathcal{F}\left[m_j \right](\mathbf{k}_l)$, is a discrete Fourier-like transformation. The operator $\mathcal{A}_{i} = \mathcal{P}_{i}\mathcal{F}^\mathrm{d}$ is illustrated in Figure~\ref{fig:Operator}. 

The continuous-to-discrete forward model \eqref{eq:ContinousOperator} becomes discrete, i.e.,  
\begin{equation}\label{eq:discrforwardmodel}
	Y_i =  \mathcal{A}_i X_i + \epsilon_i \, ,
\end{equation}	
and the corresponding discrete inverse problem is:
\begin{align}
	& \text{find} \quad X_i \in \mathbb{R}^N \quad  \text{s.t.}\nonumber\\   
	&  Y_i \approx \mathcal{A}_{i}X_i \quad \forall i    \, .           \label{eq:discrinvprob} 
\end{align}

In this paper, each measurement $Y_i$ is obtained with Cartesian undersampling.
As mentioned above, undersampling is a common practice as it reduces scanning time. However, it also reduces the amount of measurements available,  resulting in the non-uniqueness of the solution to the inverse problem and provoking aliasing artifacts \cite{subsampling}.

%% file: content/method.tex
\section{Proposed Method}\label{sec:method}

The reconstruction of $T_{2}^{*}$-mapping is usually carried out by undertaking two tasks.

The first step is to reconstruct the multicontrast image $\{X_i\} = \{X_i,\, i=1,\dots,n_\mathrm{t}\}$ by iterative reconstruction, i.e., by iteratively minimizing a functional of the form
\begin{equation}\label{eq:MBIRbasic}
    \min_{\left\{ X_{i}\right\} } \, \sum_{i=1}^{n_\mathrm{t}}\mathcal{L}_{i}(X_{i})+\beta\mathcal{R}\left(X_{1},\ldots,X_{n_\mathrm{t}}\right)
\end{equation}
where $\mathcal{L}_{i}$ represents the data fidelity term quantifying the disparity between the reconstruction $X_{i}$ and the measurement $Y_i$ through the forward operator $\mathcal{A}_i$, typically
\begin{equation}
    \mathcal{L}_i\left(X_i\right)  =\frac{1}{2}\left\|\mathcal{A}_{i}X_i-Y_{i}\right\|^{2}_{2}\label{eq:fidelityterm} 
\end{equation}
where $\|\cdot\|_{2}$ denotes the $\ell^2$-norm. $\mathcal{R}$ denotes a regularization term that imposes individual and/or joint properties on the images, and $\beta>0$ is the penalty parameter. The main contribution of this paper 
is the improvement of this first step through the design of $\mathcal{R}$.

The second step corresponds to estimating $T_2^*(\vec{r})$ at each pixel $r_j$, with $j=1,\dots,N$, by curve fitting using the magnitude of a discrete version of \eqref{eq:ModelT2} evaluated at $t=t_i$, $i=1,\dots,n_\mathrm{t}$ (see \cite{classicT1T2}). 
Namely, considering the discrete representations
\begin{equation}\label{eq:discreteT2}
    T_{2}^{*} (\vec{r})= \sum_{j=1}^{N}\tau_j m_j(\vec{r})
\end{equation}
and
\begin{equation}\label{eq:discreteC}
    \left|M_0(\vec{r})\mathrm{e}^{-i\gamma B_{0}(\vec{r})}\right|= \sum_{j=1}^{N}d_j m_j(\vec{r}) \, ,
\end{equation}
$\tau_j, d_j$ are estimated for each $j=1,\dots,N$ from equation \eqref{eq:ModelT2} evaluated at $r_j$, i.e., by solving
\begin{equation}\label{eq:modelcurvefit}
	\left|x_j(t_i)\right|\approx d_j\exp\left(-\frac{t_{i}}{\tau_j}\right),\quad  i=1,...,n_\mathrm{t}.
\end{equation}

\subsection{Conventional Methods\label{subsec:Classic-methods}}

Iterative reconstruction methods with different regularization terms have been proposed over the last decades to incorporate prior information in undersampled MRI reconstruction. Two state-of-the-art methods that exploit redundancies across the contrast dimension and \ac{JTV} and \ac{LR} reconstruction. 

\subsubsection{Joint Total Variation} 

\Ac{CS} techniques, in an \emph{analysis} model, address under-sampling by leveraging the sparsity of the reconstructed images using penalties of the form
\begin{equation}
    \mathcal{R}_\mathrm{CS}\left(\left\{ X_{i}\right\} \right)= \| T \left(\left\{ X_{i}\right\} \right) \|_q  \label{eq:cs}
\end{equation}
where $T\colon \mathbb{R}^{N \times n_\mathrm{t}} \to \mathbb{R}^M$ is a sparsifying transform and $\|\cdot\|_q$ is either the $\ell_{1}$-norm or the $\ell_{0}$ semi-norm. Examples in the literature include  \ac{TV} \cite{TVsingle,GeneralizationTV} as well as $l_1$-wavelet
regularization\cite{2009wavelet-basic,MultipleDL}. These approaches could be used to define the regularization term in \eqref{eq:MBIRbasic} by applying the penalty for each contrast independently. 

Inter-contrast information can also be exploited to ``consolidate'' the images, for example with \ac{JTV} defined as
\begin{equation}
    \mathcal{R}_\mathrm{JTV}\left(\left\{ X_i\right\} \right)=\sum_{j=1}^{N}\sqrt{\sum_{i=1}^{n_\mathrm{t}}\left\Vert \left[\nabla X_{i}\right]_{j}\right\Vert ^{2}_{2}}\label{eq:JTVReg}
\end{equation}
where $\nabla:\mathbb{R}^N\rightarrow\left(\mathbb{R}^{N}\right)^{2}$ denotes the discrete \ac{2D} gradient \cite{JTVcontrasts,JTVmethod}, which promotes joint sparsity of the $X_{i}$s across contrasts. Other examples include \ac{TNV} \cite{likeJTV-TGV} and \ac{PLS} \cite{joint-PET-MRI}. These works report noise and artifact reduction as compared to reconstructing the contrast images individually. 

\subsubsection{Nuclear Low Rank}

\Ac{LR} method involves controlling the rank of the multicontrast (Casorati) matrix achieved by promoting linear dependencies between contrast \cite{CS-MRI-1,LRmethod}. Given the shared structures across different contrasts, assuming a form of low rank in the multicontrast matrix is relevant due to the redundant information between them. The corresponding penalty is defined as
\begin{equation}
	\mathcal{R}_\mathrm{NLR}(X)=\left\Vert X\right\Vert_{*}\label{eq:LRreg}
\end{equation}
where $X = [X_1,\dots,X_{n_{\mathrm{t}}}] \in \mathbb{R}^{N\times n_{\mathrm{t}}}$, $\left\Vert \cdot\right\Vert _{*}$ is the nuclear norm defined as the sum of singular values, that is, from the singular value decomposition  $X=U\varSigma V^{T}$,  $\left\Vert X\right\Vert _{*}=\text{tr}(\varSigma)$. 

\subsection{Our Approach: CConnect}

Classical regularization methods discussed in subsection \ref{subsec:Classic-methods}, such as \ac{LR} and \ac{JTV}, enforce geometric similarities between contrasts. However, these strictly geometrical assumptions may not universally hold true, as anomalies or organ defects can modify individual patient measurements. Relying solely on prescribed similarity assumptions might fail to capture the variations inherent in patient-specific data accurately. On the contrary, \ac{ML} and more particularly \ac{DL} techniques aim at learning image characteristics from a training dataset which can be used as a prior information during image reconstruction. Our approach is derived from Uconnect for multi-energy \ac{CT} \cite{Alexpaper}, which is inspired from \ac{DiL} and described in the next subsection. 

\subsubsection{Inspiration: Multicontrast Dictionary Learning}

The general idea is to assume that the image can be sparsely decomposed in a basis of trained atoms that form a dictionary. A single decomposition for an entire image is often impractical and therefore it is achieved on a patch basis. In the context of multi-contrast \ac{MRI},  the corresponding penalty is defined as
\begin{align}
    \mathcal{R}_\mathrm{DiL}\left(\left\{ X_{i}\right\} \right) {} = {} &  \min_{\{z_p\} }\,\sum_{i=1}^{n_\mathrm{t}}\sum_{p=1}^{n_{\mathrm{p}}}\frac{1}{2}\|P_{p}X_{i}-D_{i}z_{p}\|_{2}^{2}\nonumber \\
    & {} + \alpha\|z_{p}\|_q\label{eq:DicLearning}
\end{align}
where $P_{p}\in\mathbb{R}^{d\times N}$ is the $p$-th patch extractor, $p=1,\dots,n_\mathrm{p}$ $D_{i}\in\mathbb{R}^{d\times S}$ is the dictionary that corresponds to the $i$th contrast, and each $z_{p}\in\mathbb{R}^{S}$ is the sparse code for each patch $p$. $\|\cdot\|_{q}$ ($\ell^0$ of $\ell^1$), and $\alpha>0$. \Ac{DiL} is a \emph{synthesis} model for \ac{CS}.  Note that in this formulation a single sparse $z_p$ code is used such that the information is conveyed across the contrasts $i = 1,\dots, n_\mathrm{t}$. This constraint can be relaxed by utilizing one sparse code per image with a joint sparsity norm \cite{perelli2022multi}. 

Unsupervised training of the dictionaries involves minimizing the expression \eqref{eq:DicLearning} across a training dataset containing multicontrast images. However, accurately representing all potential contrast images demands a substantial number of atoms, leading to a high computational cost during training. Additionally, the inefficiency of patch-based \ac{DiL} may arise from the shift-variant nature of atoms, resulting in duplicates during the training process. Moreover, employing numerous neighboring or overlapping patches across training images is suboptimal for sparse representation, as sparsification occurs independently on each patch.

\subsubsection{Proposed Approach}

Our method, namely CConnect, proposes an alternative to multicontrast \ac{DiL}  that does not require a patch decomposition and, more importantly, that is easier to train. Specifically, our method requires only one mapping, taking the form of an image-to-image \ac{CNN}, for each contrast. 

We define the following penalty that resembles   \eqref{eq:DicLearning},
\begin{equation}
    \mathcal{R}_\mathrm{CC}( \{ X_i\})=\min_{z \in \mathbb{R}^{N}} \,  \sum_{k=1}^{n_\mathrm{t}}\left\| f_i(z)-X_i\right\|^{2}_{2}  +  \alpha \mathcal{H} (z)\label{eq:Ourreg}
\end{equation}
where each $f_i \colon \mathbb{R}^{N} \to \mathbb{R}^N$ is a trained \ac{CNN} that maps a single latent image $z \in \mathbb{R}^N$ to an image in the $i$th channel and $\mathcal{H}$ is a regularization term for $z$ (with weight $\alpha>0$).  $\mathcal{R}_\mathrm{CC}$ is minimized if the  $X_i$s are ``connected'' through a single $z$ that is ``smooth'' in the sense of $\mathcal{H}$. 
In \cite{Alexpaper}, in the context of multi-energy spectral \ac{CT} reconstruction, $z$ corresponds to a ``reference image'' corresponding to the lowest-energy image. By connecting all the channels to a reference image, the entire raw data set $\{Y_i\}$ is used for the reconstruction of all channel $X_i$.

In this work, the latent variable takes the form of a mean image $\overline{X}$,
\begin{equation}\label{eq:MeanReal}
	\overline{X}=    \frac{1}{n_{\mathrm{t}}}\sum_{i=1}^{n_{\mathrm{t}}} X_{i}   \, , %\frac{X_{1}+\ldots+X_{n_\mathrm{t}}}{n_\mathrm{t}} \, ,
\end{equation} 
and  the \acp{CNN} $f_i$ are trained such that 
\begin{equation}\label{eq:trainbehaviour}
	f_i\left( \overline{X} \right) \approx  X_i    \quad \forall i =1,\dots,N \, .
\end{equation}
The choice of $z$ being the mean image is deliberate. Each contrast holds distinct information that is not present in the others, primarily due to the undersampling method employed. Additionally, the first contrasts have more prominent features, making extracting features from them more notorious than the other contrasts.

Finally, we used an \ac{WR} penalty $\mathcal{H} = \mathcal{R}_\mathrm{CS}$  as in \eqref{eq:cs} where $T$ was chosen as Haar wavelet transform \cite{wavelettype}, although other regularizers can be used, for example the Huber penalty as proposed in \cite{Alexpaper}. In our case, the wavelet regularization has the best result since the referent image  $z$ is not necessarily smooth.   

\subsubsection{Architecture and training \label{subsubsec:training}}

\begin{figure}[h]	
	\centering{}\includegraphics[scale=0.87]{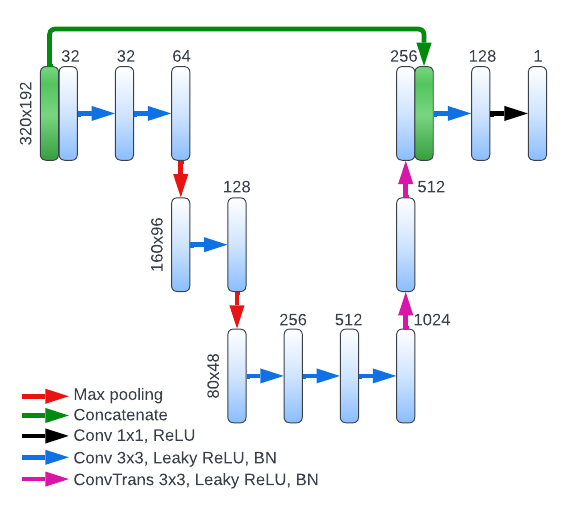}
	\caption{CConnect architecture (i.e., the $f_i$s). `Conv' denotes the discrete convolution, `ConvTrans' is the inverse operation of the convolution and BN means batch normalization. Additionally, each numerical value in the figure corresponds to the number of features in the output. The max pooling operation reduces the size of the images to half while the ConvTrans operation increases their size by two.} \label{fig:Color} 
\end{figure}

We designed CConnect's architecture (i.e., the $f_i$s) as a \ac{CNN} with an encoder and decoder comprising nine convolutional layers for a total of 215 million parameters, see Figure~\ref{fig:Color}. CConnect architecture resembles the UNet model \cite{UNet}, however the fact that it has a unique skip connection and employs leaky ReLU as activation functions make the model able to address coherent aliasing artifacts created by Cartesian undersampling, which a standard UNet is not capable to correct according to our exploratory experiments.

The \acp{CNN} $f_i$, $i=1,\dots,n_\mathrm{t}$, through the regularization \eqref{eq:Ourreg}, are required  (i) to convey information between contrasts by connecting all contrast to a single latent image $z$ and  (ii) to remove aliasing artifacts due to undersampling. For this purpose, we will augment the datasets as follows: Let $\left\{X_{k,i}^\star,\,i=1,\dots,n_\mathrm{t} \right\}^{K}_{k=1}$ be the initial dataset comprising $K$ clean contrast images sequences. For each $k=1,\dots,K$ we define the ``clean mean image" as 
\begin{equation}\label{meanclean}
	\overline{X}_k^\star  = \frac{1}{n_{\mathrm{t}}} \sum_{i=1}^{n_\mathrm{t}} X^\star_{k,i} \, .
\end{equation}
and the ``mean aliased image'' as
\begin{equation}\label{meanali}
	\widehat{X}_k = \frac{1}{n_{\mathrm{t}}} \sum_{i=1}^{n_\mathrm{t}} \widehat{X}_{k,i} \, ,
\end{equation}
where the contrasts sequence $\{\widehat{X}_{k,i}\}_{i=1}^{n_\mathrm{t}}$ are reconstructed images obtained by solving  \eqref{eq:MBIRbasic} with $\beta=0$ (using a \ac{CG} method with stopping criteria being the norm of the gradient being less than $10^{-5}$), thus suffering from remaining aliasing due to the undersampling. For each \ac{CNN} $f_i$, the training dataset is partitioned as 
\begin{equation} 
    \left\{ \left(\overline{X}^\star_k,X^\star_{k,i}\right) \right\}_{k=1}^{K}\cup
    \left\{ \left(\widehat{X}_k,X^\star_{k,i}\right) \right\}_{k=1}^{K},
\end{equation}
and each $f_i$ is trained such that 
\begin{equation}\label{eq:trainfunctionbehaviour}
	f_i\left( \overline{X}^\star_k \right) \approx  X^\star_{k,i} \quad \text{and} \quad f_i\left( \widehat{X}_k \right) \approx  X^\star_{k,i}  \quad  \forall k,i \, .
\end{equation}
The training is achieved by minimizing a mean square error as the loss function using Adam with weight decay regularization and learning rate of $10^{-4}$ \cite{adamw}. Each sequence of contrasts, $\{X^\star_{k,i}\}$ and $\{\widehat{X}_{k,i}\}$, in the dataset are normalized by the maximum value of the first clean contrast $X_{k,1}^{\star}$, $k=1,\dots,K$. 
Additionally, it is crucial to note that the trained functions depended on the specific aliasing present on $\widehat{X}_{k,i}$, which in turn depends entirely on the undersampling pattern. Consequently, it is necessary to train specific CConnect functions for each distinct undersampling pattern. 

\begin{figure}[h]
    \setkeys{Gin}{width=\linewidth}
    \centering
	\subfigure[$\overline{X}^\star$]{
		\includegraphics[width=0.3\linewidth]{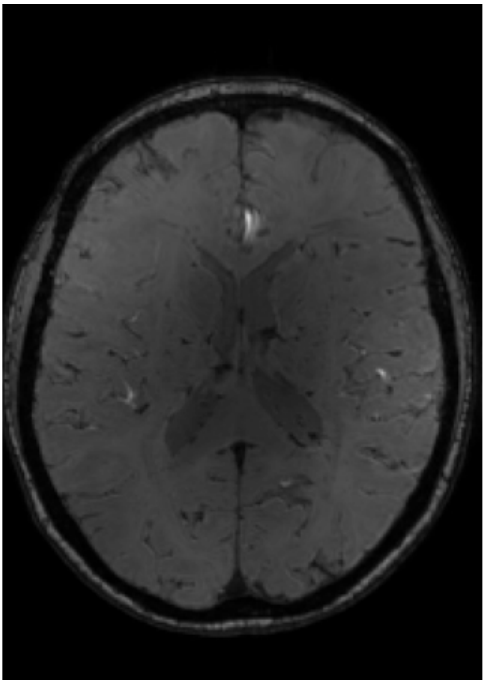}
	}%
	\subfigure[$\widehat{X}$]{
		\includegraphics[width=0.3\linewidth]{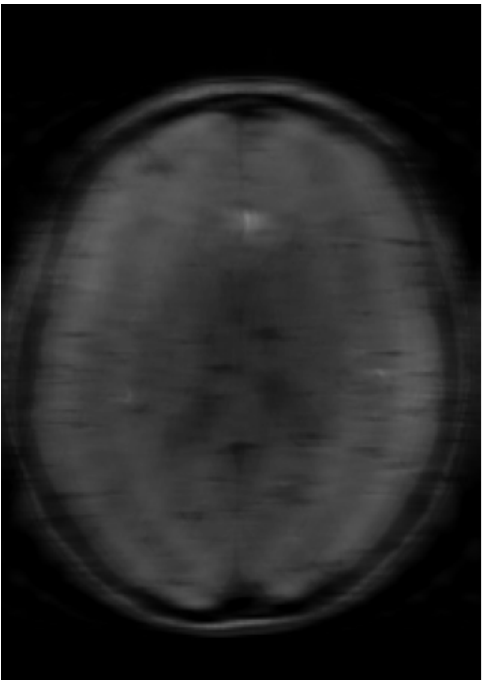}
	}
	%%%%%%%%%%%%%%%%%%%%%%%%%%%%%%%%%%%%%%%%%%%%%%%%%%%%%%%%%%%%%%%%%%%

	\subfigure[$f_{i}\left(\overline{X}^\star\right)$]{
		\includegraphics[width=0.3\linewidth]{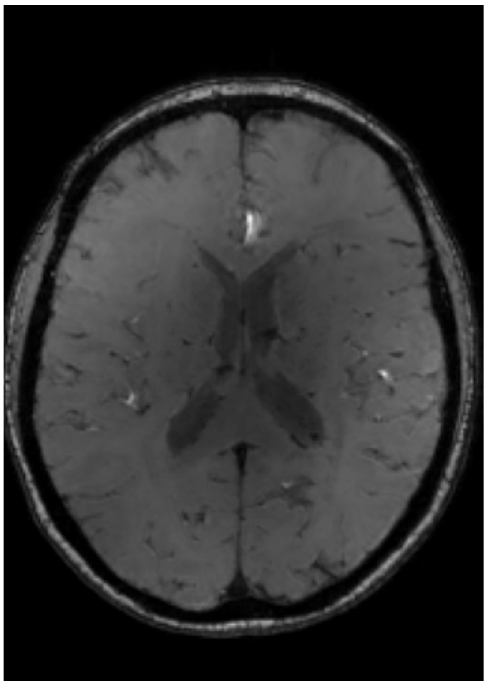}
	}%
	\subfigure[$f_{i}\left(\widehat{X}\right)$]{
		\includegraphics[width=0.3\linewidth]{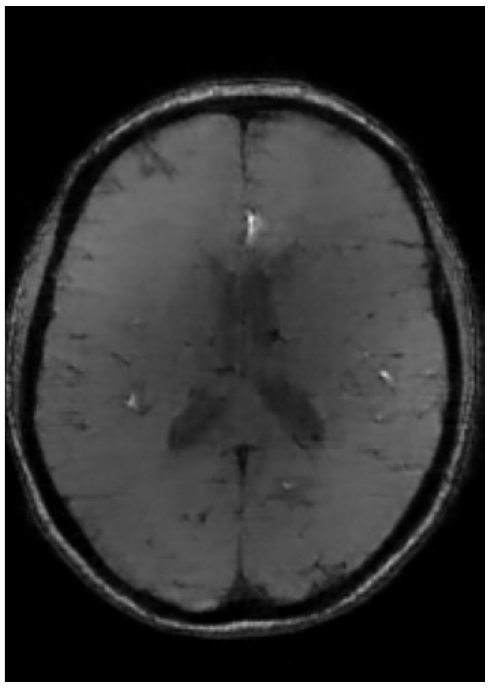}
	}%
	\subfigure[$X^\star_{i}$]{
		\includegraphics[width=0.3\linewidth]{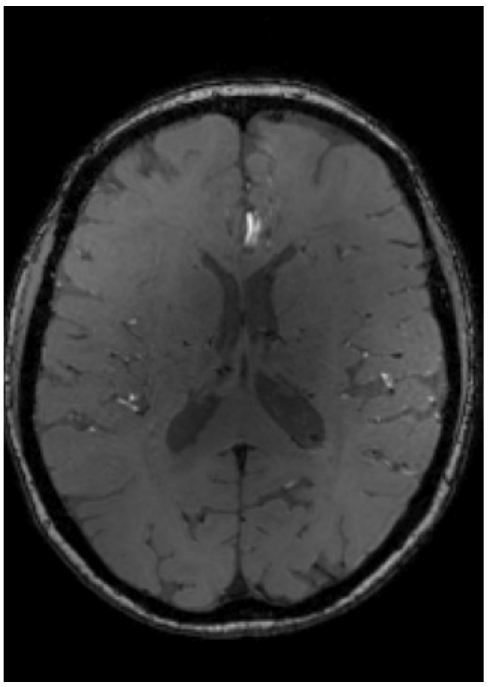}
	}%

	\caption{Illustration of a trained mapping $f_i$ ($i=3$) on an clean averaged image $\overline{X}=\frac{1}{n_{\mathrm{t}}}\sum_{i=1}^{n_{\mathrm{t}}} X_i^\star$ and a mean aliased image $\widehat{X}=\frac{1}{n_{\mathrm{t}}}\sum_{i=1}^{n_{\mathrm{t}}}\widehat{X}_{i}$.}\label{fig:TesttrainFunc}

\end{figure}

Figure~\ref{fig:TesttrainFunc} shows an example of a trained \ac{CNN} $f_i$, $i=3$, mapping a mean clean image $\overline{X}^\star = \frac{1}{n_{\mathrm{t}}}\sum_{i=1}^{n_{\mathrm{t}}} X_{i}^\star$ and a mean aliased image $\widehat{X}=\frac{1}{n_{\mathrm{t}}}\sum_{i=1}^{n_{\mathrm{t}}} \widehat{X}_i$ to the target image $X^\star$ (all of which were taken from the testing dataset). We observe that the mapped images $f_{i}\left(\overline{X}^\star\right)$ and $f_{i}\left(\widehat{X}\right)$ are similar to the target clean image $X^\star_{i}$, which shows that our trained model can generalize to a testing dataset.

\subsubsection{Reconstruction Algorithm}

Solving \eqref{eq:MBIRbasic} using the synergistic penalty \eqref{eq:Ourreg} is achieved through alternating minimization with respect to $\{X_{i}\}$ and $z$. Given the estimates $\left\{ X_{i}^{(n)}\right\}$ and $z^{(n)}$ at iteration $n$, the updated estimates at iteration $n+1$ are computed as follows: 
\begin{align}
	X_{i}^{(n+1)} = & {} \argmin_{X_{i}\in\mathbb{R}^{N}} \, L_{i}(X_{i}) +\beta\frac{1}{2}\left\Vert mf_{i}\left(\frac{z^{(n)}}{m}\right)-X_{i}\right\Vert _{2}^{2} \, \forall i \label{eq:Update_x}, \\
 	z^{(n+1)}     \in & {} \argmin_{z\in\mathbb{R}^{N}}\, \sum_{i=1}^{n_\mathrm{t}}\frac{1}{2}\left\Vert mf_{i}\left(\frac{z}{m}\right)-X_{i}^{(n)}\right\Vert_{2}^{2}	+\alpha \mathcal{H}(z)\label{eq:Update_z}\, 
   \end{align}
(where $m$ is described below). Each sub-problem is approximately solved using iterative algorithms. We used one iteration of the \ac{CG} method \cite{CGintroduction} for  \eqref{eq:Update_x}, and one outer iteration of the \ac{ADMM} \cite{boyd2011distributed} for each step on \eqref{eq:Update_z}. The initial images are defined as $X_{i}^{(0)}= \widehat{X}_i $ where $\widehat{X}_i$ is reconstructed from $Y_i$ by solving \eqref{eq:MBIRbasic} with $\beta=0$ (also with a \ac{CG} method), while $z$ was initialized as $z^{(0)} = \frac{1}{n_\mathrm{t}} \sum_{i=1}^{n_\mathrm{t}}\widehat{X}_i$. 
The stopping criterion is determined when the $\ell^2$-norm of the gradient of the sub-problem \eqref{eq:Update_x} begins to increase.

The normalization constant $m$ in \eqref{eq:Update_x} and \eqref{eq:Update_z} is chosen as $m=\max\left(X^{(0)}_1\right)$ and is required since the $f_i$ are trained with normalized sets.  

The optimization problem is not convex due to the $f_i$s and therefore proving the convergence is challenging.  Nevertheless, we observed in our experiments that the algorithm decreases the objective function at each iteration, and that it behaves as if it converges, for a wide range of $\beta$-values.

\subsection{Comparison with other DL Techniques }\label{subsec:Anothers-machine-learning}

\Ac{ML} and \ac{DL} has significantly influenced medical image reconstruction \cite{survey-ML-Reconstrution}. Several approaches share similarities with our methodology, utilizing \ac{DL} as a tool to enhance iterative reconstructions \cite{ML-Constrast-MRI,ML-iterative-CS}. However, a comparison of our proposed approach to other \ac{DL} techniques presents some challenges. For once, \ac{DL} models are usually not open-source or demand substantial efforts for replication, involving adjustment to the specific data structure and fine-tuning of the hyperparameters that are heavily dependent on the used dataset. Additionally, we are using a Cartesian undersampling patterns with a structure across contrasts that complement information through them. This hinders comparison with \ac{DL} techniques developed for other undersampling patterns that would not work well in our case \cite{DL-constrasttoT2-MANTIS}.

Alternatively, we could compare to methods that employ \ac{DL} for image post-processing, such as denoising. But for denoising techniques to achieve good results in \ac{MR} images reconstructed from undersampled measurements, the undersampling must follow radial or random patterns, since in such cases, artifacts resemble noise \cite{SurveyDenoising}, while our undersampling pattern generates strong structured aliasing artifacts. 

Nonetheless, as part of our validation, we compared our approach with a standard de-aliasing method designed by us. We label this technique \acf{DeAli} and we describe it here. The \ac{DeAli} method consists in obtaining a aliased initial image $\widehat{X}_{i}$, reconstructed from each undersampled measurement $Y_i$ as 
\begin{equation}
	\widehat{X}_i\in\argmin_{X\in\mathbb{R}^{N}} \frac{1}{2}\left\|\mathcal{A}_{i}X-Y_{i}\right\|^{2}_{2}\label{eq:noreg}
\end{equation}
using \ac{CG}, and then, the \ac{DeAli} reconstruction, is obtain by applying an image post-processing $g_i$, $i=1,\dots,n_{\mathrm{t}}$, which is a trained image-to-image \acp{CNN} such that
\begin{equation}
	g_i(\widehat{X}_i) \approx X^\star_i \, .
\end{equation}

Here, the \ac{DeAli} \acp{CNN} $g_i$ share the same design as the  CConnect  \acp{CNN} $f_i$. The unique difference lies in the definition of the training set, which is given by: 
\begin{equation}
   \left\{ \left(\widehat{X}_{k,i},X^\star_{k,i}\right) \right\}_{k=1}^{K}.
\end{equation}
for $i=1,\dots,n_\mathrm{t}$. Once the $g_i$ are trained, they are used to de-aliase any reconstruction with the same undersampling pattern. 

%% file: content/experiments.tex
\section{Experiments}\label{sec:experiments}

\subsection{Data Sets and \ac{CNN} Training}\label{dataset-training}

Multi-contrast \ac{MR} images from the SENIOR cohort \cite{datasetref} were acquired on a 7-Tesla MAGNETOM Terra scanner (Siemens Healthineers, Erlangen, Germany) equipped with a 32-channel head coil, at Biomaps, Service Hospitalier Fr\'ed\'eric Joliot, Orsay, France. 

This initial dataset had echo times $t_i$ 1.68, 4.73, 7.78, 10.83, 13.88, 16.93, 19.98, 23.03, 26.08 and 29.13~ms, i.e., 10 contrasts, a repetition time of 37~ms, voxel size of 0.8~mm $\times$0.8~mm$\times$0.8~mm, a 256~mm field-of-view, 740 Hz/Px bandwidth and were reconstructed using GRAPPA \cite{SENSE-GRAPPA}. The total acquisition time was 9 minutes and 48 seconds per patient. The complete dataset included five patients, and each of them contained the magnitude of 10 $T_2^*$-weighted \ac{3D} brain images of dimensions 192$\times$320$\times$320. 

From the initial dataset, we extracted 100 (out of 320) high-quality (i.e., low-noise, aliasing-free) multi-contrast brain \ac{2D} \ac{MR} images (slices) from each patient.
The $K=300$ multi-contrast images from the first three patients were used to form the \acp{CNN} training set. We formed the 
development set of early stopping criteria in the \acp{CNN} training with the multi-contrast images from the 4th patient, while we used the multi-contrast images from the 5th patient to evaluate the proposed reconstruction method.

Each high-quality multi-contrast (real-valued) image $\{X_i^{\star},\, i =1,\dots,n_{\mathrm{t}}  \}$ was used as a \ac{GT} reference image. To simulate the raw data and retrospective undersampling $\{Y_i,\, i =1,\dots,n_{\mathrm{t}}\}$ we used the discrete forward model \eqref{eq:discrforwardmodel} (with $\epsilon_i=0$), i.e., the simulated measurement $Y_i$ was obtained by performing the Fourier transform of $X_i^\star$ and then undersampling the result with the sampling operator $\mathcal{P}_i$. 

\begin{figure}[h]
	\centering
	\subfigure[Factor 4]{
		\includegraphics[width=0.3\linewidth]{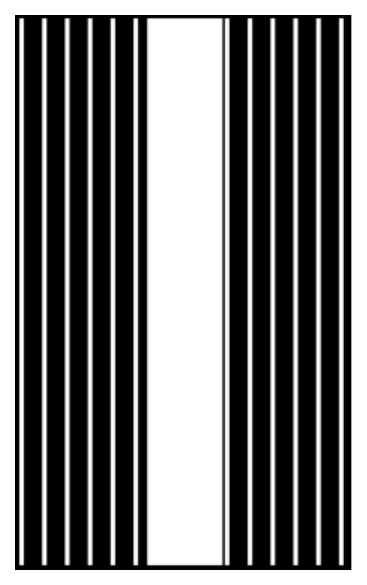}
	}%
	\subfigure[Factor 5]{
		\includegraphics[width=0.3\linewidth]{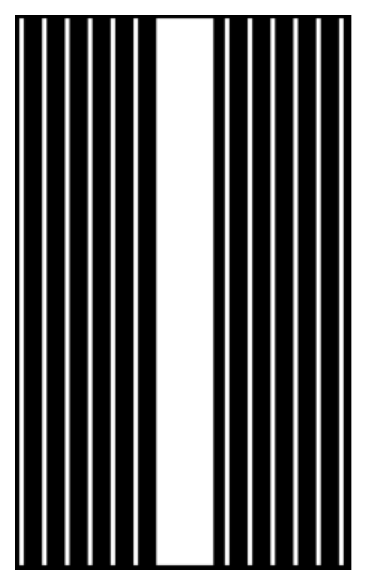}
	}%
	\subfigure[Factor 6]{
		\includegraphics[width=0.3\linewidth]{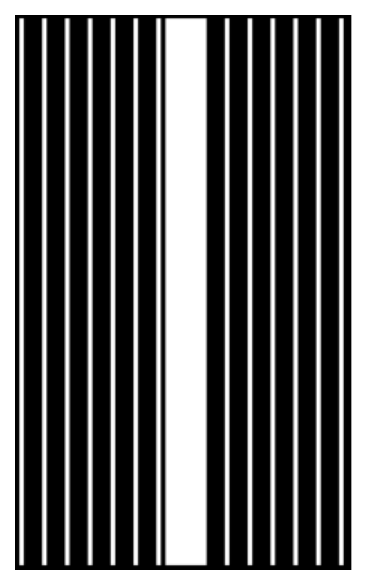}
	}%
	\\  
	%\medskip
	
	\subfigure[$R=6$ pattern $\quad$ associated to $Y_3$]{
		\includegraphics[width=0.3\linewidth]{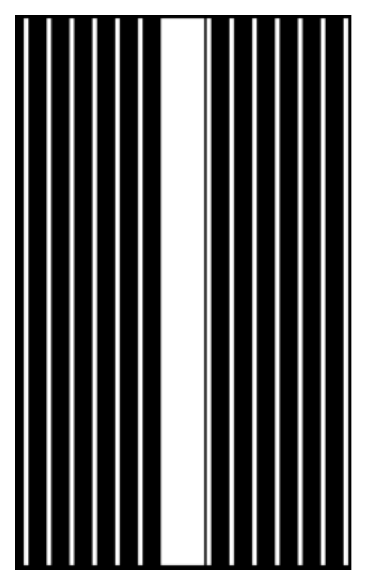}
	}%
	\subfigure[$R=6$ pattern $\quad$ associated to $Y_8$]{
		\includegraphics[width=0.3\linewidth]{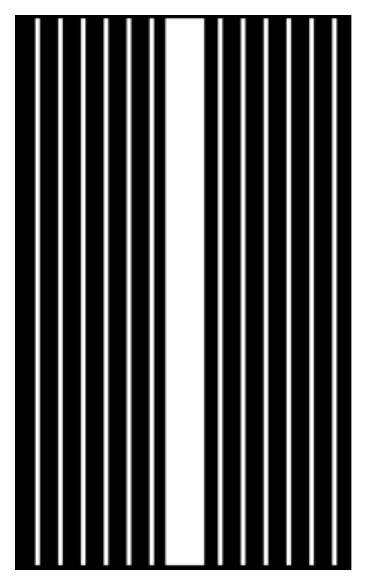}
	}%
	\subfigure[Overlap of $Y_3$ and $Y_8$ subsampling patterns]{
		\includegraphics[width=0.3\linewidth]{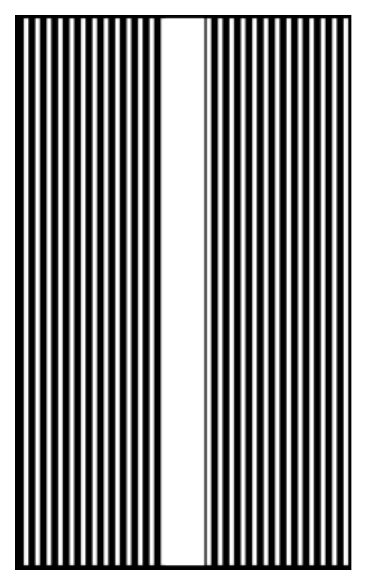}
	}%
	
	\caption{Illustration of the undersampling pattern for three acceleration factors. The first row shows the undersampling patterns for acceleration factors $R=4$, $R=5$ and $R=6$ associated with measurement $Y_1$. The second row shows the undersampled pattern for an acceleration factor $R=6$ for the third and eighth measurements, and their overlap.}\label{fig:Cartesian-undersampling}
\end{figure}

We considered three different undersampling factors, $R=4$, $R=5$, and $R=6$, using a uniform Cartesian trajectory with a fully sampled central region. For $R=4$ the pattern fully samples 32 readouts in the k-space center and uniformly samples 16 readouts in the outer region of the k-space. Similarly, for $R=5$ the pattern samples 24 readouts in the k-space center and 8 readouts in the region of the k-space, whereas for $R=6$ the pattern samples 16 readouts in the k-space center and uniformly samples 8 readouts in the outer region of the k-space. For a fixed factor $R$, the undersampling patterns $\mathcal{P}_i$ for different contrasts $i$ are such that they all fully sample the k-space center, but the outer sampled readouts are shifted and do not overlap between contrasts (see Figure~\ref{fig:Cartesian-undersampling}). 

\input{./curves/psnr_ssim}

\begin{figure*}[h]
	\begin{overpic}[width=\linewidth, tics=5]{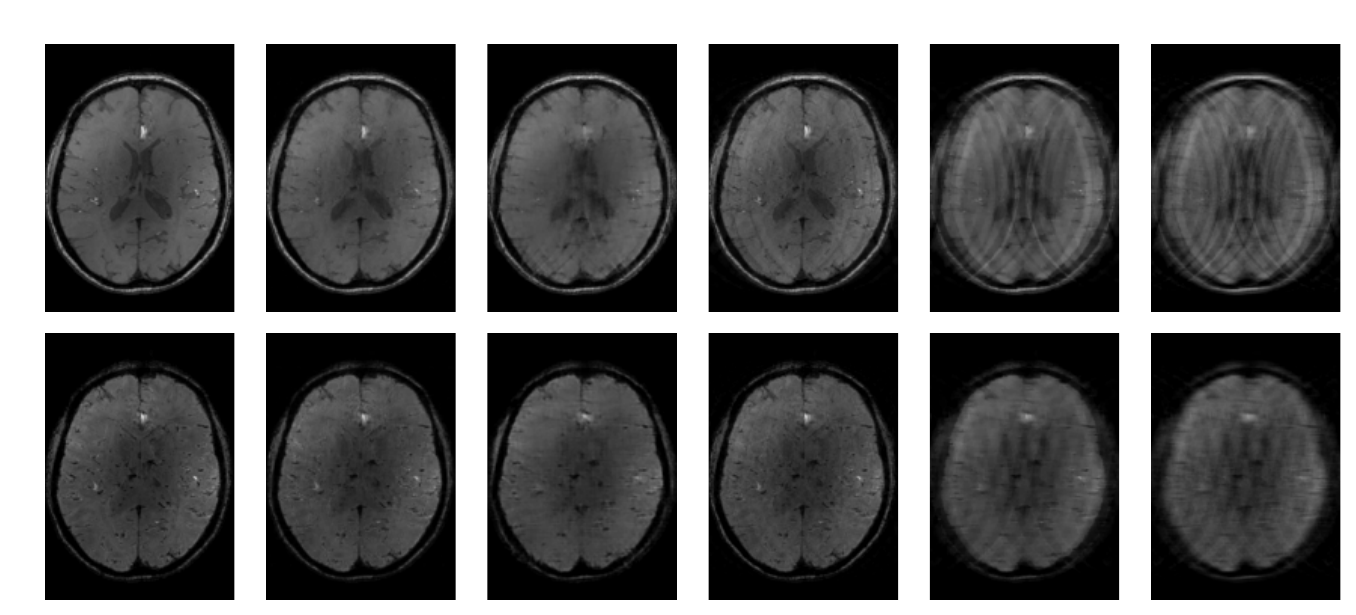}
		\put (0,31) {$X_3$}
		\put (0,9) {$X_7$}
		\put (9,43) {GT}
		\put (22.7,43) {CConnect}
		\put (40.3,43) {\ac{DeAli}}
		\put (57.3,43) {\ac{LR}}
		\put (74,43) {\ac{JTV}}
		\put (84.5,43) {No Regularization}
	\end{overpic}
	\vspace*{-5mm}
	\centering
	\caption{Reconstruction results for undersampling factor $R=6$ for slice 70 of the test set. GT, the proposed CConnect, DeAli, NLR, JTV and no regularization are included.}\label{fig:Ej70-f6}
\end{figure*}

Specialized software and high-performance hardware are crucial elements in \ac{DL}.  We used  TensorFlow and JAX \cite{jax2018github} to implement and train the \acp{CNN}, specifically utilizing specialized JAX libraries such as Optax and Flax, for efficient implementation and fine-tuning of the neural networks. In terms of hardware,  we used an NVIDIA GeForce GTX 1660 Ti with 6GB memory.

\subsection{Metrics}

In order to evaluate the quality of a reconstructed image with respect to a reference image, we used \ac{SSIM} and \ac{PSNR} \cite{metrics}. These metrics are important in evaluating and quantifying image fidelity and perceptual quality. Each metric offers unique insights into different aspects of image quality, ranging from pixel-level differences to human perceptual sensitivity. \Ac{PSNR} tends to focus more on pixel-wise differences and is easier to compute, while \ac{SSIM} considers structural information and human perception, making it more robust for evaluating perceptual quality. We utilized the functions \texttt{structural\_similarity} and \texttt{peak\_signal\_noise\_ratio} from the Python package \texttt{skimage} to compute the \ac{PSNR} and \ac{SSIM}.

\input{./tables/table1}

\input{./tables/table2}

\subsection{Results}

\subsubsection{$T_{2}^{*}$-weighted Image Reconstruction}

 Reconstructions results for $T_{2}^{*}$-weighted images using no regularization (naive-solution), \ac{DeAli}, JTV, NLR and the proposed CConnect method are shown in Figure~\ref{fig:Ej70-f6} for $R=6$ in a representative 2D slice in comparison to the GT. The penalty parameters were fixed for each reconstruction. The values are $\beta=5\cdot10^6$ for \ac{LR}, $\beta=5\cdot10^6$ for \ac{JTV} and finally, $\beta=10^4$ and $\alpha=10^6$ for CConnect, which are chosen for each method in their optimal range. Significant remaining aliasing is observed for the No Reg and JTV methods. DeAli and NLR achieved better results, although remaining aliasing is still visible for the 3th contrast in this case. CConnect achieves the best results for both contrasts.

Table~\ref{tab:mean50slide} shows the mean and standard deviation of metrics across the reconstruction of $T_{2}^{*}$-weighted images from 50 distinct \ac{2D} slices in the testing dataset, considering undersampling factors of  $R=4$, $R=5$, and $R=6$.  Notably, CConnect emerges with the most favorable mean values, and the standard deviation exhibits a comparable order of magnitude across the evaluated methods. The computation time for all the methods is in the same order of magnitude (a few hours). CConnect is the most time-intensive, taking 2 to 3 times longer than \ac{JTV} or \ac{LR}, and \ac{DeAli} is the fastest, being a non-iterative method.

Finally, to analyze the methods' robustness concerning the penalty parameter $\beta$, we present the evaluation metrics of \ac{JTV}, \ac{LR}, and CConnect methods for different values of $\beta$ on all factors in the case slice 50 (Figure~\ref{fig:betasej50final}). The range of $\beta$ is displayed in a relative scale for each method, in order to compare robustness between the different methods. CConnect and \ac{LR} exhibit similar \ac{PSNR} values for different penalty parameters $\beta$; nevertheless, our method outperforms \ac{LR} in the \ac{SSIM} metric. Additionally, all methods demonstrate a reasonable degree of stability for a wide range of values of $\beta$.

\subsubsection{$T_{2}^{*}$ map Reconstruction}

After the reconstruction of the ten contrast images, the final step is recovering the $T_{2}^{*}$ map. For this, we use the 2-parameter model shown in Equation \eqref{eq:modelcurvefit} for each pixel of the image. Here, the values of $t_{i}$ are described in subsection \ref{dataset-training}, and we use the contrasts obtained by the different methods. The metrics are calculated over masked reconstructions that exclude regions of non-physical $T_{2}^{*}$ values (pixels with or adjacent to values of 170~ms or more) to ensure a more meaningful comparison.

\begin{figure*}
	\begin{overpic}[width=\linewidth, tics=5]{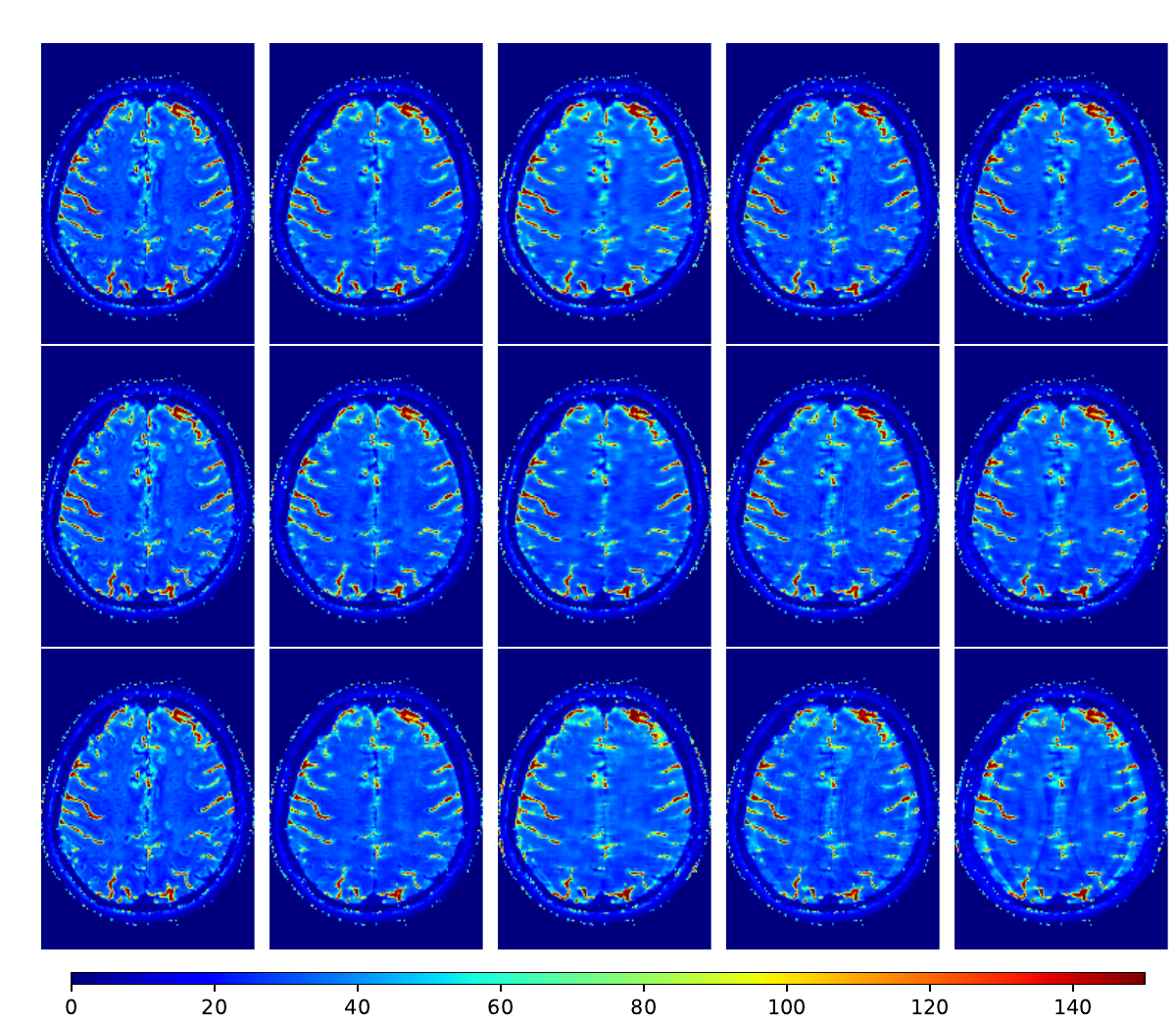}
		\put (1.5,67) {\rotatebox{90}{Factor 4}}
		\put (1.5,41) {\rotatebox{90}{Factor 5}}
		\put (1.5,15) {\rotatebox{90}{Factor 6}}
		\put (11,84) {GT}
		\put (28,84) {CConnect}
		\put (49,84) {\ac{DeAli}}
		\put (68.8,84) {\ac{LR}}
		\put (88.5,84) {\ac{JTV}}
	\end{overpic}
	\vspace*{-5mm}
	\caption{\label{fig:T2ej50} $T_2^*$ maps for the slice 50 for the GT and different reconstructions methods CConnect, DeAli, NLR and JTV for undersampling factors of $R=4$, $R=5$ and $R=6$.}
\end{figure*}

\begin{figure*}
    \begin{overpic}[width=\linewidth, tics=5]{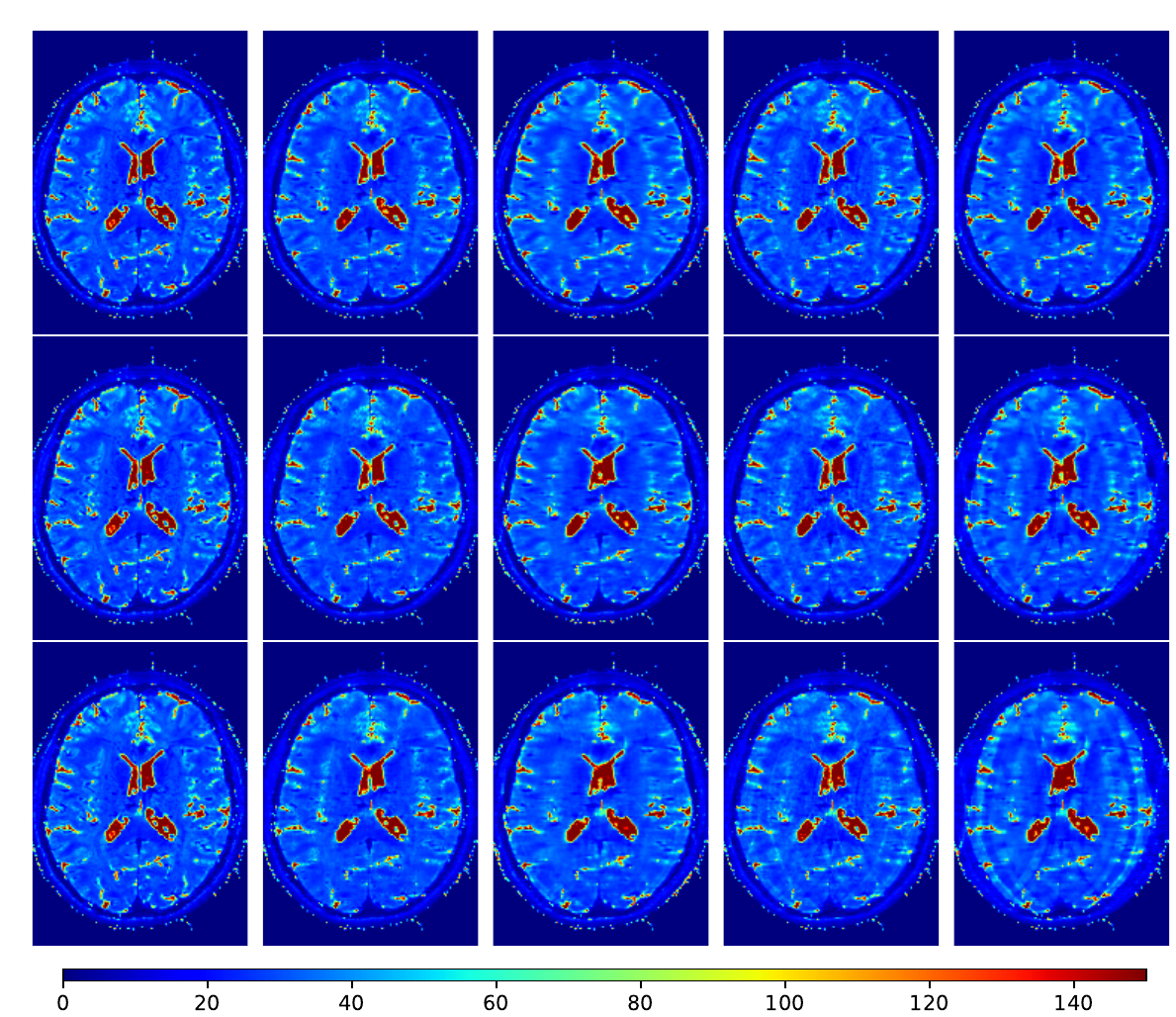}
	    \put (0.5,67.5) {\rotatebox{90}{Factor 4}}
	    \put (0.5,42) {\rotatebox{90}{Factor 5}}
	    \put (0.5,15) {\rotatebox{90}{Factor 6}}
	    \put (10.5,84.5) {GT}
	    \put (28,84.5) {CConnect}
	    \put (49,84.5) {\ac{DeAli}}
	    \put (69,84.5) {\ac{LR}}
	    \put (88.7,84.5) {\ac{JTV}}
    \end{overpic}
	\vspace*{-5mm}
	\caption{\label{fig:T2ej70} $T_2^*$map for the slice 70 for the GT and reconstructions methods CConnect, DeAli, NLR and JTV for undersampling factors of $R=4$, $R=5$ and $R=6$.}
\end{figure*}

Figure~\ref{fig:T2ej50} and Figure~\ref{fig:T2ej70} show the reconstructed $T_{2}^{*}$ maps for all the methods, at the slices 50 and 70 correspondingly (of the 5th patient), with undersampling factors of $R=4$, $R=5$, and $R=6$. Table~\ref{tab:T2metrics} presents the statistical results for the three undersampling factors across 50 slices from the test set.

In the experiments, CConnect consistently outperforms the \ac{DeAli} and the classic methods in all metrics. Furthermore, both CConnect and \ac{DeAli} demonstrated superior capability in eliminating aliasing artifacts compared to conventional methods. However, \ac{DeAli} seems to remove some image details.  

%% file: curves/psnr_ssim.tex
\begin{figure}
	\centering
	\ref{named2} \\
	\subfigure[PSNR factor 4]{
		\begin{minipage}[t]{0.47\linewidth}
			\begin{tikzpicture}
				\begin{axis}[
					height=0.75\linewidth,
					width=1.2\linewidth,
					tick label style={font=\scriptsize},
					x label style={at={(axis description cs:0.5,0.2)},font=\scriptsize},
					xlabel={$\beta/\beta_\mathrm{max}$},
					legend columns=-1,%the legend are plotted horizontally
					legend entries={CConnect, \ac{JTV},\ac{LR}},
					legend to name=named2,% stored in named
					]
					\addplot[color=blue, mark=otimes*, mark size=1.5pt] table[x=beta/max_beta, y=PSNR, col sep=comma] {./data/pnsr-fac4-our.txt};
					\addplot[color=orange, mark=diamond*, mark size=1.5pt] table[x=beta/max_beta, y=PSNR, col sep=comma] {./data/pnsr-fac4-jtv.txt};
					\addplot[color=red, mark=diamond*, mark size=1.5pt] table[x=beta/max_beta, y=PSNR, col sep=comma] {./data/pnsr-fac4-lr.txt};
				\end{axis}
			\end{tikzpicture}
		\end{minipage}%
	}%
	\subfigure[SSIM factor 4]{
		\begin{minipage}[t]{0.47\linewidth}
			\begin{tikzpicture}
				\begin{axis}[
					height=0.75\linewidth,
					width=1.2\linewidth,
					tick label style={font=\scriptsize},
					x label style={at={(axis description cs:0.5,0.2)},font=\scriptsize},
					xlabel={$\beta/\beta_\mathrm{max}$},
					]
					\addplot[color=blue, mark=otimes*, mark size=1.5pt]       table[x=beta/max_beta, y=SSIM, col sep=comma]    {./data/ssim-fac4-our.txt};
					\addplot[color=orange, mark=diamond*, mark size=1.5pt] table[x=beta/max_beta, y=SSIM, col sep=comma] {./data/ssim-fac4-jtv.txt};
					\addplot[color=red, mark=diamond*, mark size=1.5pt] table[x=beta/max_beta, y=SSIM, col sep=comma] {./data/ssim-fac4-lr.txt};
				\end{axis}
			\end{tikzpicture}
		\end{minipage}%
	}
	\subfigure[PSNR factor 5]{
		\begin{minipage}[t]{0.47\linewidth}
			\begin{tikzpicture}
				\begin{axis}[
					height=0.75\linewidth,
					width=1.2\linewidth,
					tick label style={font=\scriptsize},
					x label style={at={(axis description cs:0.5,0.2)},font=\scriptsize},
					y label style={at={(axis description cs:0.2,0.5)},anchor=south,font=\scriptsize},
					xlabel={$\beta/\beta_\mathrm{max}$},
					]
					\addplot[color=blue, mark=otimes*, mark size=1.5pt] table[x=beta/max_beta, y=PSNR, col sep=comma] {./data/pnsr-fac5-our.txt};
					\addplot[color=orange, mark=diamond*, mark size=1.5pt] table[x=beta/max_beta, y=PSNR, col sep=comma] {./data/pnsr-fac5-jtv.txt};
					\addplot[color=red, mark=diamond*, mark size=1.5pt] table[x=beta/max_beta, y=PSNR, col sep=comma] {./data/pnsr-fac5-lr.txt};
				\end{axis}
			\end{tikzpicture}
		\end{minipage}%
	}%
	\subfigure[SSIM factor 5]{
		\begin{minipage}[t]{0.47\linewidth}
			\begin{tikzpicture}
				\begin{axis}[
					height=0.75\linewidth,
					width=1.2\linewidth,
					tick label style={font=\scriptsize},
					x label style={at={(axis description cs:0.5,0.2)},font=\scriptsize},
					y label style={at={(axis description cs:0.2,0.5)},anchor=south,font=\scriptsize},
					xlabel={$\beta/\beta_\mathrm{max}$},
					]
					\addplot[color=blue, mark=otimes*, mark size=1.5pt]       table[x=beta/max_beta, y=SSIM, col sep=comma]           {./data/ssim-fac5-our.txt};
					\addplot[color=orange, mark=diamond*, mark size=1.5pt] table[x=beta/max_beta, y=SSIM, col sep=comma] {./data/ssim-fac5-jtv.txt};
					\addplot[color=red, mark=diamond*, mark size=1.5pt] table[x=beta/max_beta, y=SSIM, col sep=comma] {./data/ssim-fac5-lr.txt};
				\end{axis}
			\end{tikzpicture}
		\end{minipage}%
	}
	\subfigure[PSNR factor 6]{
		\begin{minipage}[t]{0.47\linewidth}
			\begin{tikzpicture}
				\begin{axis}[
					height=0.75\linewidth,
					width=1.2\linewidth,
					tick label style={font=\scriptsize},
					x label style={at={(axis description cs:0.5,0.2)},font=\scriptsize},
					y label style={at={(axis description cs:0.2,0.5)},anchor=south,font=\scriptsize},
					xlabel={$\beta/\beta_\mathrm{max}$},
					]
					\addplot[color=blue, mark=otimes*, mark size=1.5pt] table[x=beta/max_beta, y=PSNR, col sep=comma] {./data/pnsr-fac6-our.txt};
					\addplot[color=orange, mark=diamond*, mark size=1.5pt] table[x=beta/max_beta, y=PSNR, col sep=comma] {./data/pnsr-fac6-jtv.txt};
					\addplot[color=red, mark=diamond*, mark size=1.5pt] table[x=beta/max_beta, y=PSNR, col sep=comma] {./data/pnsr-fac6-lr.txt};
				\end{axis}
			\end{tikzpicture}
		\end{minipage}%
	}%
	\subfigure[SSIM factor 6]{
		\begin{minipage}[t]{0.47\linewidth}
			\begin{tikzpicture}
				\begin{axis}[
					height=0.75\linewidth,
					width=1.2\linewidth,
					tick label style={font=\scriptsize},
					x label style={at={(axis description cs:0.5,0.2)},font=\scriptsize},
					y label style={at={(axis description cs:0.2,0.5)},anchor=south,font=\scriptsize},
					xlabel={$\beta/\beta_\mathrm{max}$},
					]
					\addplot[color=blue, mark=otimes*, mark size=1.5pt] table[x=beta/max_beta, y=SSIM, col sep=comma]{./data/ssim-fac6-our.txt};
					\addplot[color=orange, mark=diamond*, mark size=1.5pt] table[x=beta/max_beta, y=SSIM, col sep=comma] {./data/ssim-fac6-jtv.txt};
					\addplot[color=red, mark=diamond*, mark size=1.5pt] table[x=beta/max_beta, y=SSIM, col sep=comma] {./data/ssim-fac6-lr.txt};
				\end{axis}
			\end{tikzpicture}
		\end{minipage}%
	}%
	\caption{PSNR and SSIM mean values of the reconstructed $T_2^*$-weighted images using different values of penalty parameter $\beta$ for the iterative methods.}\label{fig:betasej50final}
\end{figure}

%% file: tables/table1.tex
\begin{table*}
	\centering
	\caption{Metrics of the reconstruction of $T_{2}^{*}$-weighted images, average computed for the reconstruction of slides 30 to 79.}\label{tab:mean50slide}
	\begin{tabular}{c|cc|cc|cc|}
		\cline{2-7}
		& \multicolumn{2}{c|}{$R=4$}     & \multicolumn{2}{c|}{$R=5$}    & \multicolumn{2}{c|}{$R=6$}           \\ \cline{2-7} 
		& \multicolumn{1}{c|}{PNSR}    &  SSIM   & \multicolumn{1}{c|}{PNSR}    &  SSIM  & \multicolumn{1}{c|}{PNSR}    &   SSIM   \\ \hline
		\multicolumn{1}{|c|}{DeAli} & \multicolumn{1}{c|}{$29.191\pm1.345$}    &  0.869  $\pm$ 0.019   & \multicolumn{1}{c|}{26.802 $\pm$ 1.28}    &  0.815 $\pm$ 0.019  & \multicolumn{1}{c|}{25.06 $\pm$ 0.947}    &   0.757 $\pm$ 0.021   \\ \hline
		\multicolumn{1}{|c|}{JTV}   & \multicolumn{1}{c|}{30.84 $\pm$ 1.281}    &  0.898 $\pm$ 0.015   & \multicolumn{1}{c|}{28.283 $\pm$ 0.979}    &   0.833 $\pm$ 0.021  & \multicolumn{1}{c|}{25.8 $\pm$ 0.987}    &   0.749 $\pm$ 0.02   \\ \hline
		\multicolumn{1}{|c|}{NLR}      & \multicolumn{1}{c|}{\textbf{33.038 $\pm$ 2.195}}    &  0.905 $\pm$ 0.027   & \multicolumn{1}{c|}{31.828 $\pm$ 2.176}    &  0.876 $\pm$ 0.033  & \multicolumn{1}{c|}{30.722 $\pm$ 2.082}    &   0.85 $\pm$ 0.035   \\ \hline
		\multicolumn{1}{|c|}{\textbf{CConnect}}      & \multicolumn{1}{c|}{32.834 $\pm$ 2.064} & \textbf{0.922 $\pm$ 0.022} & \multicolumn{1}{c|}{\textbf{32.108 $\pm$ 2.125}} & \textbf{0.907$\pm$ 0.028} & \multicolumn{1}{c|}{\textbf{31.55 $\pm$ 2.094}} & \textbf{0.895 $\pm$ 0.03} \\ \hline
	\end{tabular}
\end{table*}

%% file: tables/table2.tex
\begin{table*}
	\centering
	\caption{Metrics of the reconstruction of $T_{2}^{*}$ map, average computed for the reconstruction of slides 30 to 79.}\label{tab:T2metrics} 
	\begin{tabular}{c|cc|cc|cc|}
		\cline{2-7}
		& \multicolumn{2}{c|}{$R=4$}     & \multicolumn{2}{c|}{$R=5$}    & \multicolumn{2}{c|}{$R=6$}           \\ \cline{2-7} 
		& \multicolumn{1}{c|}{PNSR}    &  SSIM   & \multicolumn{1}{c|}{PNSR}    &  SSIM  & \multicolumn{1}{c|}{PNSR}    &   SSIM   \\ \hline
		\multicolumn{1}{|c|}{DeAli}      & \multicolumn{1}{c|}{28.52$\pm$0.95}    &  0.902 $\pm$ 0.015   & \multicolumn{1}{c|}{28.896 $\pm$ 0.787}    &  0.887 $\pm$ 0.021  & \multicolumn{1}{c|}{26.44 $\pm$ 0.504}    &   0.849 $\pm$ 0.022   \\ \hline
		\multicolumn{1}{|c|}{JTV} & \multicolumn{1}{c|}{30.507 $\pm$ 0.779}    &  0.918 $\pm$ 0.009   & \multicolumn{1}{c|}{29.07 $\pm$ 0.544}    &  0.883 $\pm$ 0.019  & \multicolumn{1}{c|}{27.289 $\pm$ 0.45}    &   0.85 $\pm$ 0.021   \\ \hline
		\multicolumn{1}{|c|}{NLR}   & \multicolumn{1}{c|}{30.162 $\pm$ 0.523}    &  0.912 $\pm$ 0.014   & \multicolumn{1}{c|}{29.39 $\pm$ 0.542}    &   0.895 $\pm$ 0.017  & \multicolumn{1}{c|}{28.995 $\pm$ 0.506}    &   0.879 $\pm$ 0.017   \\ \hline
		\multicolumn{1}{|c|}{\textbf{CConnect}}      & \multicolumn{1}{c|}{\textbf{30.801 $\pm$ 0.576}} & \textbf{0.92 $\pm$ 0.011} & \multicolumn{1}{c|}{\textbf{30.018 $\pm$ 0.802}} & \textbf{0.907$\pm$ 0.016} & \multicolumn{1}{c|}{\textbf{29.213 $\pm$ 0.729}} & \textbf{0.894 $\pm$ 0.017} \\ \hline
	\end{tabular}
\end{table*}

%% file: content/discussion.tex
\section{Discussion}\label{sec:discussion}

We introduced a novel iterative reconstruction for multi-contract \ac{MRI} reconstruction, utilizing trained \acp{CNN} in the penalty term. CConnect showed superior performance compared to classical multi-contrast reconstruction methods and a deep learning approach, in terms of visual image quality and \ac{PSNR} and \ac{SSIM} metrics. Notably, as the undersampling factor increases, our method outperforms the other methods more noticeably. The CConnect architecture employed in our trained functions, when trained appropriately, is able to remove aliasing artifacts induced by Cartesian undersampling. The utilization of a latent image, playing the role of a mean across contrasts, proved to be crucial as it encapsulates information from all contrasts and undersampling patterns. 

The mappings $\left\{ f_{i}\right\} $ demonstrated their efficacy by (i) sharing redundant information among contrasts and (ii) effectively eliminating aliasing resulting from undersampling measurements. However, these \acp{CNN} are designed for specific undersampling patterns and fixed echo times; therefore, measurements with different trajectories and mapping sequences will require retraining of the proposed CConnect method. 

We must point out that the trained $\{f_i\}$ \acp{CNN} can be enhanced. For instance, we could potentially improve the model by jointly training the $\{f_i\}$ as a single \ac{CNN} with $n_{t}$ outputs. This architecture could incorporate additional physical aspects of the reconstruction process, such as echo times or direct input of K-space measurements. Additionally, other improvements can be obtained as proposed in \cite{ML-InverseProblemgeneral,ML-Constrast-MRI,ML-iterative-CS}, where iterative \ac{DL} models remove artifacts and noise progressively through multiple layered evaluations.

Further improvements in the results could be achieved by fine-tuning all hyperparameters, which was not the primary objective of this paper. Also, weights in the regularizing term, corresponding to a confidence estimation of the trained function $\{f_i\}$, might improve the method, since empirical evidence indicates that all functions do not exhibit similar behavior. Notably, the $f_i$ functions corresponding to intermediate contrasts performs better than functions corresponding to the initial and final contrasts. This observation aligns with our expectations, considering that the mean of the contrasts is more closely related to the central contrasts.

A general comment is that improvement in the reconstruction of the contrast images does not automatically translate into an improved reconstruction of the $T_2^*$ maps. Indeed, a large improvement in the reconstruction of the contrast images can become a negligible improvement in the reconstruction of the $T_2^*$ map. This is exemplified in our experiments in the case of sampling factor $R=4$, where CConnect outperforms \ac{JTV} in contrast reconstruction (see Table~\ref{tab:mean50slide}) but has similar metrics in $T_2^*$  map reconstruction (see Table~\ref{tab:T2metrics}). Also, it is important to point out that even if CConnect outperforms the other methods as we increase the subsampling factor $R$, there is a limit on how much we can undersample and still obtain clinically meaningful $T_2^*$ map reconstructions. The two aspects just mentioned above are inherent challenges of this technique.

In addition, it would be of interest to compare the reconstruction obtained with the proposed CConnect method to reconstruction obtained with other classic methods, such as \ac{WR} or switching from the $\ell_1$-norm to the $\ell_{0}$ semi-norm. Likewise, the comparisons could encompass other \ac{ML} techniques, although this presents several challenges as mentioned in Section~\ref{subsec:Anothers-machine-learning}.

Our study presents several limitations. CConnect was evaluated with retrospective undersampling simulated from magnitude contrast-weighted images, considering a single coil-acquisition. Future work should extend the proposed approach to multi-coil complex value reconstruction from prospectively undersampled k-space data. Furthermore, the proposed approach was tested on data from a single subject. Further testing in a larger cohort of healthy subjects and patients should be investigated as future work. Despite data being 3D, reconstruction was performed in 2D due the limited number of datasets available for training. Extending the proposed reconstruction to 3D will be also of interest. Additionally, the proposed approach was evaluated for $T_2^*$ mapping, however the framework should be easily extended to other parametric maps and will be investigated in the future. 

%% file: content/conclusion.tex
\section{Conclusion}\label{sec:conclusion}

In this study, we developed a synergistic regularization term leveraging trained \acp{CNN}. Drawing inspiration from the dictionary method \cite{DLearning} and the methodology presented in \cite{Alexpaper} for reconstructing \ac{CT} images across multiple energies, our approach demonstrated notable improvements in $T_{2}^{*}$ map reconstructions of the brain compared to state-of-the-art methods.  This methodology may be extended to other \ac{qMRI} parameters as future work. While there are wide options for improvement, our approach may mark a significant step forward in utilizing \ac{DL} for multicontrast reconstruction.